\documentstyle[twoside,epsf]{article}

\input ibvs2.sty

\begin{document}

\IBVShead{6108}{28 May 2014}

\IBVStitle{Cepheids and RR Lyrae stars in the K2 fields}

\IBVSauth{Moln\'ar, L.$^{1,2}$, Plachy, E.$^{1,2}$, Szab\'o, R.$^1$}

\IBVSinst{Konkoly Observatory, Research Centre for Astronomy and Earth Sciences, Konkoly Thege Mik\'os \'ut 15-17, H-1121 Budapest, Hungary}
\IBVSinst{Institute of Mathematics, Physics, and Engineering, Savaria Campus, University of West Hungary,\linebreak K\'arolyi G\'asp\'ar t\'er 4, H-9700 Szombathely, Hungary}
\IBVSinsto{e-mail: molnar.laszlo@csfk.mta.hu}

\IBVSabs{With the step-and-stare approach of the K2 mission, Kepler will be able to observe }
\IBVSabs{a large number of Cepheid an RR Lyrae stars. In this paper we summarize }
\IBVSabs{the target selection efforts, the scientific goals that can be achieved by Kepler, }
\IBVSabs{and the first impressions based on the K2 two-wheel engineering test data.}

\begintext
\section{Introduction}
The \textit{Kepler} space telescope was launched into orbit in early 2009 with a task to observe about 170 000 stars with unprecedented photometric precision and temporal coverage (Borucki et al.\ 2010). The original field-of-view was located in the Lyra-Cygnus region at high ecliptic latitudes. \textit{Kepler} observed this field for almost exactly 4 years, providing revolutionary data for several areas, but the failure of its second reaction wheel in May 2013 effectively ended its original mission. 

The amount of classical pulsators, Cepheids and RR Lyrae stars, was somewhat limited in the Lyra-Cygnus field. Before the start of the mission, 57 stars were proposed as RR Lyrae targets but 23 of them soon turned out to be other types of stars (Kolenberg et al.\ 2010). A few targets were discovered later either by the ASAS survey or as contaminants in the pixel apertures of other stars, bringing back the sample size close to 50. 

About half of the fundamental-mode RR Lyrae stars turned out to be modulated. The analysis of the 4-year-long rectified data sets revealed that a very high fraction, about 80\% shows multiple modulation periods (Benk\H{o} et al.\ 2014). Several of the Blazhko stars show various additional modes and/or period doubling. The sample size, however, is low: Benk\H{o} et al.\ (2014) created rectified data sets for only 15 stars after excluding the blended targets and the bright and heavily saturated RR Lyr, the eponym of the class. A larger population of modulated stars is needed to determine the occurrence rates and especially the various interrelations of the dynamical effects. 

Other space photometric missions observed only a handful of RR Lyrae stars. MOST measured AQ Leo, the archetype of the double-mode subclass but most RR Lyraes are too faint for the small telescope (Gruberbauer et al.\ 2007). Thirteen stars (10 RRab, 2 RRc 1 RRd stars) were identified in the first CoRoT fields (from IR1 to LRC04) and others are expected in the later fields but a thorough search has not been carried out yet (Szab\'o et al.\ 2014).

Before the mission, only a single classical Cepheid, V1154 Cyg was known in the \textit{Kepler} field. Despite a thorough search, no other stars were discovered (Szab\'o et al. 2011). V1154 Cyg itself turned out to be a rewarding target, revealing significant fluctuations in its light curve (Derekas et al.\ 2012). Similar effects were discovered in two other classical Cepheids with MOST (Evans et al.\ 2014). A handful of stars have been classified as Cepheids in the early CoRoT data, but those have not been investigated in detail, therefore it is unclear which subtypes are present in the sample (Debosscher et al.\ 2009). 

\section{The K2 mission}
After the failure of the second reaction wheel, the operators of \textit{Kepler} invited the scientific community to submit white papers detailing the possible uses and techniques for the space telescope. With the help of this feedback, a new mission scenario was devised. Named K2, after both the two-wheel operation mode and the enigmatic and challenging peak in the Karakoram Range, the new mission will carry out shorter-duration campaigns along the ecliptic and will cover several fields in the sky. This setting will allow the telescope to balance against the radiation pressure of the Sun while maintaining attitude in the other two directions with the remaining reaction wheels. 

Although the 75-day duration of the campaigns will be much shorter than the time base of the prime mission, the availability of very diverse fields will obviously provide adequate justification for it. Moreover, the 75-day-long campaigns are still longer than most space-based observations. They were only exceeded by the CoRoT long runs that lasted for about 150 days. In the foreseeable future, the regions around the ecliptic poles will be covered by TESS for about one year and after that only PLATO, to be launched in 2024, will carry out multi-year observations (Rauer et al.\ 2014).

\subsection{Target selection}
Target selection for the first few fields, including the two-wheel engineering test run, had to be carried out on short time scales, usually within a few weeks. Surveying the fields with new observations was unfeasible, hence we only proposed stars that were already marked as (potential) Cepheid or RR Lyrae stars in various databases. 

As a start we queried the stars designated as either RR Lyrae or some subclass of Cepheids from SIMBAD. The main drawback of SIMBAD is that in most cases we cannot verify the classification with actual light curves. Therefore we searched the time-domain photometric databases thoroughly. The All Sky Automated Survey\footnote{http://www.astrouw.edu.pl/asas/} (ASAS; Pojmanski 1997) provides classifications but the photometric accuracy of the data drops considerably below 10-11th magnitudes. We also used the databases of the Northern Sky Variability Survey\footnote{http://skydot.lanl.gov/} (NSVS; Wo\'zniak et al.\ 2004), SuperWASP\footnote{http://wasp.cerit-sc.cz} (Paunzen et al.\ 2014) and the INTEGRAL Optical Monitoring Camera\footnote{http://sdc.cab.inta-csic.es/omc} (IOMC; Mas-Hesse et al.\ 2003).

The two most valuable sources for RR Lyrae targets were the asteroid-searching surveys that scanned large portions of the sky repeatedly with moderate-sized telescopes to detect faint objects. Most of these data was already mined for RR Lyrae stars to conduct statistical studies and search for stellar streams and other structures in the galactic halo. The LINEAR database contains all types of RR Lyrae stars (Sesar et al.\ 2013) but covers right ascensions between 8 an 18 hours only and mostly above the celestial equator\footnote{http://www.astro.washington.edu/users/ivezic/linear/PaperIII/PLV.html}. The Catalina Survey data\footnote{http://catalinadata.org} covers a larger area, most of the sky above $\delta \sim -20\deg $, except for a wide band along the Milky Way. A second, deeper survey (Mt.~Lemmon Survey, MLS) was carried out along the ecliptic plane as well, but in both cases only the RRab stars were identified (Drake et al.\ 2013a,b). The LONEOS Phase I data that mostly covered the ecliptic was also mined for RRab stars (Miceli et al.\ 2008). The distribution of RR Lyrae stars from the various databases are shown in Figure 1. The complete classification of the Catalina data was released only very recently and will be used for target selection from Field 4 onwards (Drake et al.\ 2014). 

\IBVSfig{10cm}{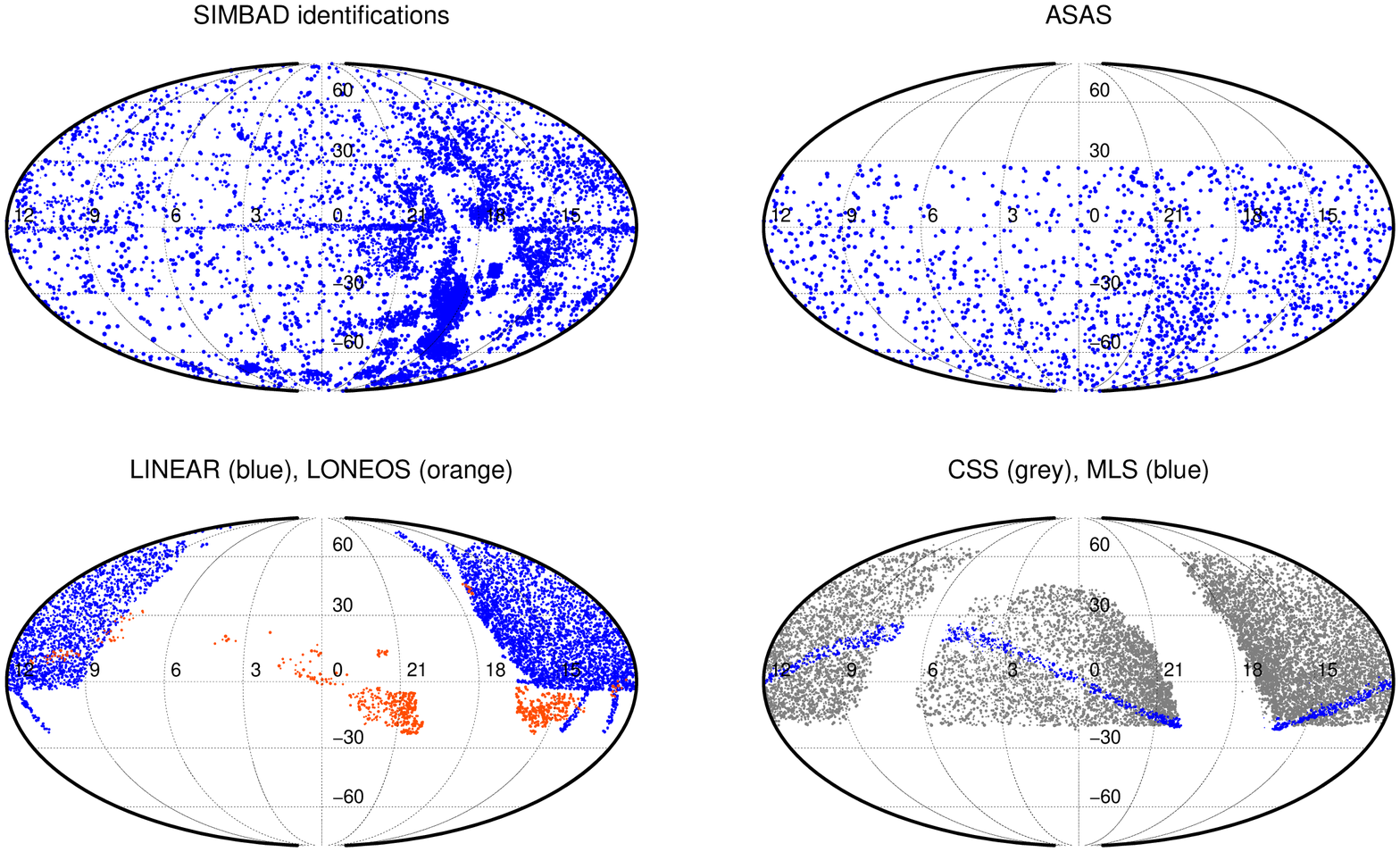}{Distribution of RR Lyrae stars in the five major data sets we used for target selection. }
\IBVSfigKey{6108-f1.eps}{sky maps}{other}

\IBVSfig{5cm}{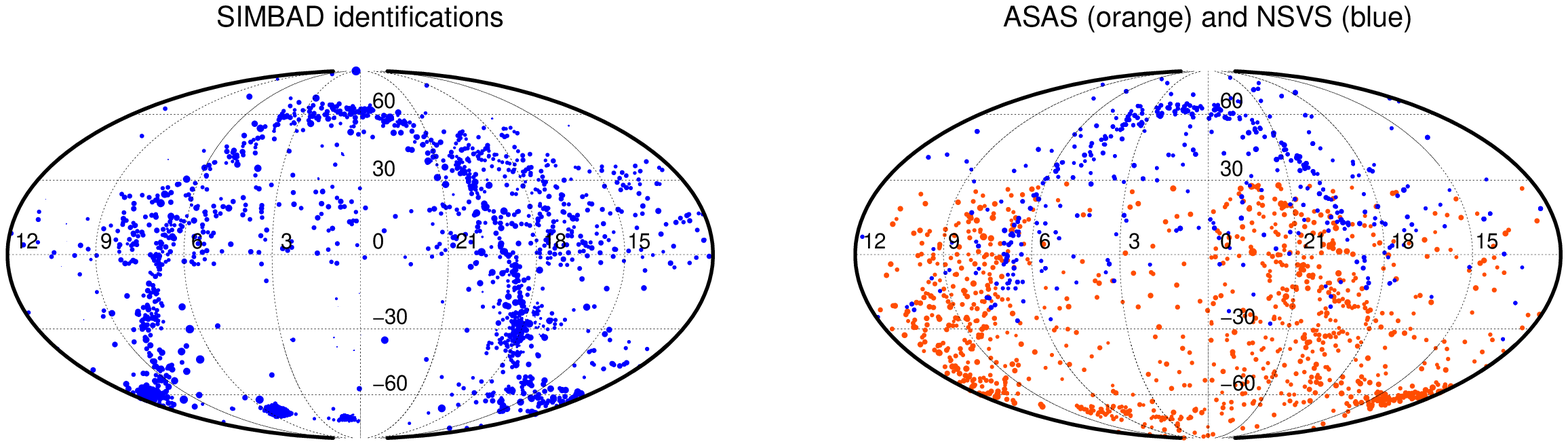}{Distribution of various Cepheid and Cepheid-like stars in SIMBAD and the ASAS and NSVS surveys. Most stars are close the plane of the Miky Way. }
\IBVSfigKey{6108-f2.eps}{sky maps}{other}

Cepheids have been classified only in SIMBAD and the ASAS and NSVS databases (Hoffmann, Harrison \& McNamara 2009; Schmidt et al.\ 2013) so we relied upon those and tried to confirm as many of them as possible with published observations. The distribution is shown in Figure 2.

\subsection{Two-wheel engineering test and Field 0}
A 9-day engineering test run was carried out in February 2014 to verify the operations with two reaction wheels. \textit{Kepler} observed some 2000 stars in the direction of Pisces. The data suffers from various pointing adjustments and drifts that make the extraction of light curves a complicated task. We found 27 RRab, 3 RRc and a possible Cepheid star among the observations, although the period of the latter star is about 53 days, much longer than the data set itself.

The brightest RRab star in the sample was ASAS J233637-0212.7 (EPIC 60018644). We extracted the light curve of the star with the PyKE toolset to test the stability and quality of the \textit{K2} data. The telescope was repositioned after BJD = 56695.359, shifting the stars by a couple of pixels on the CCDs, therefore we used two pixel masks at different positions to extract the photometric data. The resulting background-corrected light curve can be seen in Figure 3. A detailed summary of the RR Lyrae stars in the engineering-test data will be published in a future paper.

\IBVSfig{8.8cm}{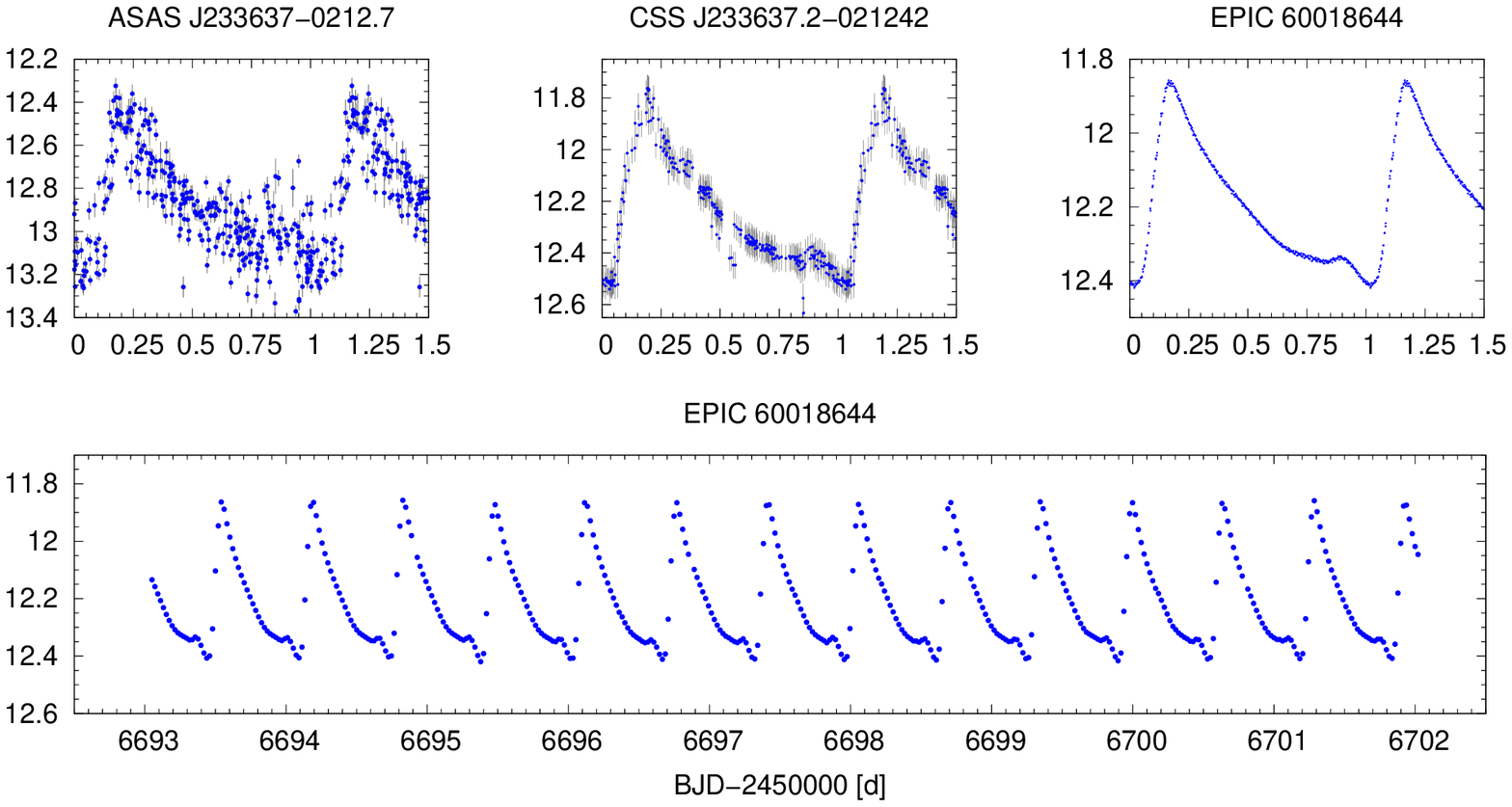}{Comparison of the ASAS, Catalina and \textit{K2} long cadence data of the RR Lyrae star ASAS J233637-0212.7. The upper panels show the light curves folded with a pulsation period of $P=0.6450427$ days. The lower panel contains the extracted \textit{K2} light curve itself.}
\IBVSfigKey{6108-f3.eps}{ASAS J233637-0212.7}{light curve}

After the engineering tests, a full-scale performance test was carried out. Campaign 0 was a shakedown that closely resembled the proposed operations of the \textit{K2} mission. Field~0 was set towards the galactic anticenter in Gemini, including the bright open cluster Messier 35. Since the asteroid searching surveys avoided this area we ended up with only 10 RR Lyrae stars that fell on silicon. However, Field 0 was ideally positioned to observe Cepheids: 14 stars were accepted to the long-cadence target list, including fundamental-mode and overtone stars, likely from both Type I and Type II Cepheids. Field 0 in itself will double the number of Cepheids observed with space-based photometry. One RR Lyrae (EW Gem; Schmidt \& Reiswig 1993) and one first-overtone Cepheid (NSVS 9770315) were selected for short cadence observations.

\subsection{Fields 1 and 3 - galore of RR Lyrae stars}
The first science field of the \textit{K2} mission was set towards the Leo-Virgo region and far from the plane of the Galaxy. With the help of the LINEAR and Catalina catalogs, we identified and proposed no less than 133 field RR Lyrae stars. This number far exceeds the approximately 50 stars that were observed in the original \textit{Kepler} field and offered the possibility for an unprecedented opportunity for statistical studies. Moreover, apart from the 118 RRab and 14 RRc stars, we found a single double-mode (RRd) star, LINEAR 2122319, in the field as well. Unfortunately, very few stars made it to the final target list, but later fields hopefully will accumulate an expected few hundred stars to carry out the statistical studies. On the other hand, six targets were selected for short cadence observations, including all three types RR Lyrae stars.

In addition, we found three intriguing extragalactic stars. The dwarf spheroidal galaxy Leo IV falls into Field 1 and three RR Lyraes were identified in it by Moretti et al.\ (2009). These stars, along with the brightest giants and supergiants in the galaxy were included in the target list. The brightness of the three RR Lyraes is below $Kp=21$ magnitudes so they represent a considerable observational challenge. During the primary mission the estimated precision for a 21st magnitude star was about 0.15 ($P=148700$ ppm) in a single long-cadence observation. The precision of the \textit{K2} measurements is expected to be within a factor of 2 in fine pointing mode (Howell et al.\ 2014), therefore the precision of individual points will be around 0.2-0.3 magnitudes. Based on that, the entire data set is expected to provide an accuracy of 5 millimagnitudes for a coherent signal. Most additional-mode peaks fall below this limit, but the strongest ones can be recovered (Benk\H{o} et al.\ 2014; Moln\'ar et al.\ 2012).

\subsection{Field 2 - globular clusters}

The speciality of Field 2 is the inclusion of two globular clusters, Messier 4 and 80. \textit{Kepler}, with its 4$\prime\prime$/px resolution was not designed to observe dense stellar fields and that leads to various consequences. M4 is fairly spread-out, with a half-light radius ($R_h$) of 65 pixels, but its bright core would saturate the CCD. Possibly for that reason, only the northern edge falls on silicon. M80 is fainter but also more compact with a half-light radius of only 4 pixels. Still, we identified many pulsating variables in the outskirts of the clusters out to about $7\,R_h$ where we expect the crowding to be acceptable. Several RR Lyrae stars, a few semiregular and SX Phe stars and a single type II Cepheid were proposed for the two clusters. 

Apart from the globular clusters, we proposed about 50 field RR Lyrae stars and a few Cepheid/W Vir candidates, as well.

\subsection{Future fields}
The approximate positions of most of the future fields contain a large number of RR Lyrae stars, including the RRc and RRd classes. Field 7 will be positioned close to the galactic plane and therefore lack asteroid survey data. Nevertheless, the GCVS contains a large amount of otherwise unobserved RR Lyrae stars in that region (Samus et al.\ 2004).

A comparison of the distribution of various Cepheid stars with the preliminary positions of the \textit{K2} fields suggests that every field will contain a handful of Cepheids (barring misidentifications in the various survey data), including all subtypes: classical Cepheids, BL Her, and W Vir stars.

We identified two galaxies where it is possible to observe extragalactic Cepheids. Fields 7 and 8 may include NGC 6822 and IC 1613, respectively. IC 1613 is closer and contains several Cepheids that are brighter than 21 magnitudes in \textit{V} band (Bernard et al.\ 2010). NGC 6822 is somewhat farther away, but about a dozen variables are brighter than 22 magnitudes and therefore we may expect reasonable photometry from those as well (Pietrzy\'nski et al.\ 2004; Mennickent et al.\ 2006). 

The initial data products of the \textit{K2} campaigns will be the target pixel files: a times series of small CCD subframes containing the image of the star. We already gained experiences with target pixel files to create the rectified RR Lyrae data sets (Benk\H{o} et al.\ 2014), therefore the reduction of the \textit{K2} photometric data will be a relatively straightforward task. An example (EPIC 60018657) from the 9-day engineering run is displayed in Fig.\ 4.

\IBVSfig{7cm}{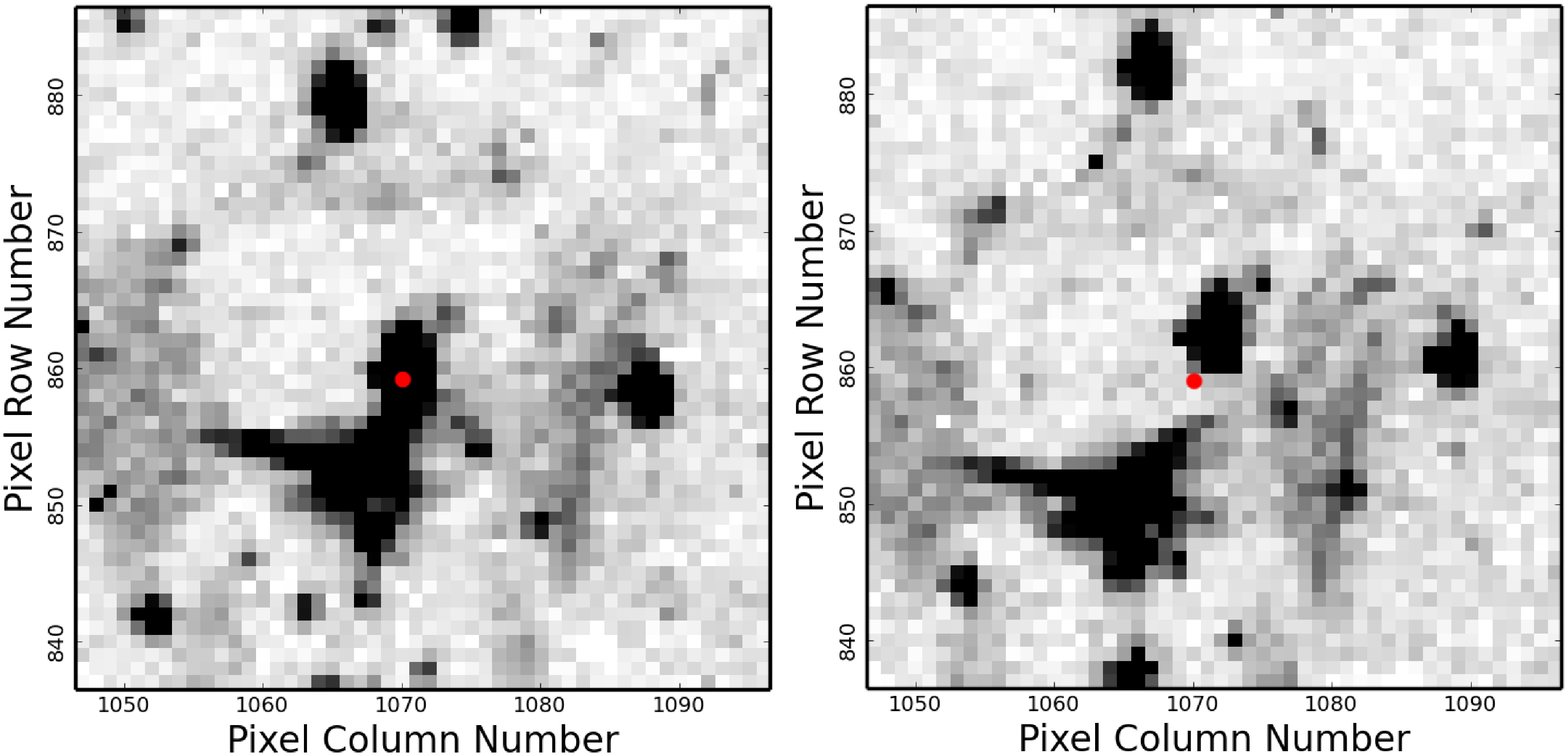}{\textit{K2} target pixel files from the 9-day engineering test run. Red dot marks the approximate photocenter of the star during the first 2 days. Left panel: 1st cadence, right panel: the 150th, i.e.\ after the repositioning of the telescope. There are also four patches of reflected light in the mask, and the brightest one contaminates the star slightly during the first part of the observations. The patches then move to the opposite direction compared to the stars, separating the bright blob from the target.}
\IBVSfigKey{6108-f4.eps}{target pixel file}{other}

\section{Scientific goals and possibilities in the \textit{K2} mission}
The step-and-stare approach of the \textit{K2} mission differs significantly from the original mission scenario. The length of the campaigns seriously limits some applications, e.g.\ the detection of long Blazhko periods and/or the variations in the modulation cycles. However, the availability of multiple fields opens up several new possibilities compared to the prime mission.

\subsection{Cepheids - pulsation and atmospheric dynamics}
Continuous observations of space photometric data revealed that classical Cepheids exhibit light curve fluctuations (Derekas et al.\ 2012). These could possibly be connected to large convective hot spots on the surface of the star (Neilson \& Ignace 2014). There are some indications that this effect is stronger in overtone stars than in fundamental-mode pulsators. If this relation exists, it may aid the determination of pulsation modes. Confirmation, however, requires several stars with different periods to be observed.

First-overtone Cepheids may turn out to be double-mode pulsators or can exhibit non-radial modes, according to the OGLE observations (Moskalik \& Ko{\l}aczkowski 2009). Some of them show strong O--C variations too, but it is not clear if that manifests in the pulsation amplitudes as well or not.

Due to their lower metallicity, light variations of type II Cepheids (W Virginis and BL Herculis stars) exhibit noticeable variations. Period doubling was already detected in one BL Her star, but hydrodynamic models predict other effects such as chaotic pulsation or low-amplitude modulation, too (Smolec et al.\ 2012; Smolec \& Moskalik 2014). Continuous, high-precision data is the best way to detect such irregularities in the pulsation of these stars.

A few anomalous Cepheids were also identified in the fields. These stars lie between the classical and type II Cepheids and follow a separate P-L relation in the period-luminosity diagram (Nemec, Nemec \& Lutz, 1994). They have low metallicities and their origin is somewhat uncertain but may involve mass transfer in a binary system with a possible link to blue stragglers (Szabados, Kiss \& Derekas 2007). \textit{K2} could be the first space telescope to observe anomalous Cepheids.

\subsection{RRab Lyrae stars -- mode interactions and the cause of the Blazkho effect}
One of the great surprises of the \textit{Kepler} mission was the detection of millimagnitude-level additional modes in almost all modulated RR Lyrae stars (Benk\H{o} et al.\ 2010, 2014). The occurrence of these modes raise serious questions about the mode selection mechanisms in RR Lyrae stars. 

Although hydrodynamic models can explain the occurrence of some of the additional modes, there is still discrepancy between the observations and the theoretical results. Based on the sample of the \textit{Kepler} field, most of the stars exhibit period doubling, related to the 9th overtone (Koll\'ath, Moln\'ar \& Szab\'o 2011), and a frequency peak that may correspond to the second overtone ($P_2/P_0 \approx 0.6$; Benk\H{o} et al.\ 2014). However, in the models, period doubling leads to the destabilization of the first overtone which was detected in very few stars. Finding more stars where the first overtone is excited can lead to accurate comparisons with the models, including the mode amplitudes and the signs of mode interactions, possibly even chaos (Plachy, Koll\'ath \& Moln\'ar 2013). Such nonlinear asteroseismic analysis was attempted only for RR Lyr itself to date (Moln\'ar et al.\ 2012). 

Mode interactions and resonances are the best candidates so far to finally explain the mysterious Blazhko effect (Buchler \& Koll\'ath 2011). Considering that only half of the RRab stars are modulated, a large survey is necessary to understand the relation between these modes. 75 days is long enough to cover at least one modulation cycle for the majority of the Blazhko stars. 

Another interesting aspect of the Blazhko effect is the apparent decline of its frequency above $P\sim 0.66-0.7$ days (Smolec 2005). It is yet unclear if the typical modulation amplitudes decrease and therefore simply become less detectable by ground-based surveys, or the mechanism of the Blazhko effect itself depends on the pulsation period, e.g.\ mode resonances may be less likely to occur. All stars in the \textit{Kepler} field have periods below 0.69~days, therefore it could not address this issue (Nemec et al.\ 2013). If the \textit{K2} campaigns can build up a suitably large sample of modulated stars including long-period ones, the observations may shed light on the origins of this effect and the inner workings of the Blazhko effect.

\subsection{RRc and RRd stars}
Modulation is present but much less common among the first-overtone (RRc) stars. Only a few RRc stars has been observed from space so far, and none of them turned out to be modulated yet: \textit{Kepler} could be the first space telescope to detect one. 

A great surprise of the original \textit{Kepler} field was that all four RRc stars turned out to be multiperiodic. Moreover, they all shared the same properties: the additional mode was detected at $P_X/P_1 = 0.60-0.64$, and showed period doubling in all cases (Moskalik 2014). We note that similar, mysterious modes were detected not only in RRc, but in RRab and LMC Cepheid stars as well, always at the same period ratio (Moskalik 2014). A closer look at the frequency tables of the two RRd stars where additional modes were reported also reveals this mode, although the authors classified them differently. Interestingly, it is connected to the first overtone in both cases (Gruberbauer et al.\ 2007; Chadid 2012). 

The origin of the $P_{X}$ (or $P_{0.6}$) modes is not yet understood. A thorough survey is required to find out whether all RRc and RRd stars exhibit it, or there is some connection between the mode selection mechanism and the physical properties of the stars. The same is true for the apparent period doubling of this mode. Luckily, the Catalina and LINEAR surveys will provide several RRc and RRd targets, especially from Field 4 onwards.

Double-mode stars are also important on their own right. The two main modes can be modeled accurately with the existing hydrodynamic codes, providing strong constraints on the physical parameters such as the metallicity and the mass of the star.

\subsection{Population studies}
Field RR Lyrae stars have very different metallities, between [Fe/H]$=-0.05$ and $-2.13$, with one outlier at $-2.54$ (Nemec et al. 2013). A change in the metal content can shift the mode resonance regions to different stellar parameters, as in the case of period doubling (Koll\'ath et al.\ 2011). If mode resonances are behind the Blazhko effect, this may lead to differences in the modulation properties as well. To detect such metallicity-dependent effects, we need to observe a large number of stars in the Galaxy. 

These investigations can be expanded further with the inclusion of stellar populations that share very similar metal content. The globular clusters M4 and M80 have distinctly different metallicities: their [Fe/H] index is --1.16 and --1.75, respectively. The dwarf galaxy Leo IV is very metal poor with [Fe/H] $= -2.4\pm 0.2$, so it can trace the low end of the metallicity sequence.

\section{Conclusions}
Although the \textit{K2} mission was born out of the unfortunate failures of the reaction wheels on board the \textit{Kepler} space telescope, the scientific potential of the new campaigns can exceed that of the original field in several areas. 

All types of Cepheid variables are rather rare in the Milky Way: only a handful is expected in every \textit{K2} field. Therefore they do not require a large pixel budget per campaign, but the step-and-stare approach can accumulate a good sample of both classical and type II stars. Cepheids also represent a great opportunity for extragalactic \textit{K2} observations in the nearby dwarf galaxies. Light-curve fluctuations and detection of additional modes can provide important insights into these stars.

RR Lyrae stars, like Cepheids, are an important step in the cosmic distance ladder, but they are good tracers of the halo structures and dwarf galaxies around the Milky Way. Therefore the understanding of their pulsations is important for the galactic structure and evolution studies. Yet several open questions still remain: the Blazhko effect, the mysterious $P_{0.6}$ mode, the role of mode interactions and the level of agreement between the observations and the 1D pulsation models. Most of the RR Lyrae stars are faint, below 14th magnitude, therefore the pixel usage would be moderate even for a high number of targets. The capabilities of \textit{Kepler} and the campaign mode of the \textit{K2} mission represent an ideal opportunity to solve these questions, provided that sufficient number of stars (preferably a few hundred in total) will be observed during the mission.

\vskip 1 cm

\emph{Acknowledgements:} The work of L. Moln\'ar leading to this research was supported by the European Union and the State of Hungary, co-financed by the European Social Fund in the framework of T\'AMOP 4.2.4.\ A/2-11-1-2012-0001 `National Excellence Program'. R.Sz.\ was supported by the J\'anos Bolyai Research Scholarship of the Hungarian Academy of Sciences. This work has been supported by the Hungarian OTKA grant K83790,  and the `Lend\"ulet-2009' Young Researchers' Programme of the Hungarian Academy of Sciences. The research leading to these results has received funding from the European Community's Seventh Framework Programme  (FP7/2007-2013) under grant agreements no. 269194 (IRSES/ASK) and no. 312844 (SPACEINN). This research has made use of the SIMBAD database, operated at CDS, Strasbourg (France), NASA's Astrophysics Data System Bibliographic Services, and PyKE (Still \& Barclay 2012), an open source software package developed and distributed by the NASA Kepler Guest Observer Office.

\references

Benk\H{o}, J.\ M., et al., 2010, \textit{MNRAS}, \textbf{409}, 1585

Benk\H{o}, J.\ M., Plachy, E., Szab\'o, R., Moln\'ar, L., Koll\'ath, Z., 2014, \textit{ApJS}, submitted

Bernard, E.\ J., et al., 2010, \textit{ApJ}, \textbf{712}, 1259

Borucki, W.\ J., et al., 2010, {\it Science}, {\bf 327}, 977

Buchler J.~R., Koll\'ath Z. 2011, \textit{ApJ}, \textbf{731}, 24

Chadid, M., 2012, \textit{A\&A}, \textbf{540}, 68

Debosscher, J., et al., 2009, \textit{A\&A}, \textbf{506}, 519

Derekas, A., et al., 2012, \textit{MNRAS}, \textbf{425}, 1312

Drake, A.\ J., et al. 2013a, \textit{ApJ}, \textbf{763}, 32 

Drake, A.\ J., et al. 2013b, \textit{ApJ}, \textbf{765}, 154

Drake, A.\ J., et al. 2014, \textit{ApJS}, accepted, arXiv:1405.4290 [astro-ph.SR]

Evans, N.\ R., Szab\'o, R., Szabados, L., Derekas, A., Kiss, L., Matthews, J., Cameron, C., 2014, \textit{IAUS}, \textbf{301}, 55

Gruberbauer, M., et al., 2007, \textit{MNRAS}, \textbf{379}, 1498

Hoffmann, D.\ I., Harrison, T.\ E., McNamara B.\ J., 2009, \textit{AJ}, \textbf{138}, 466

Howell, S.\ B., et al., 2014, \textit{PASP}, accepted, arXiv:1402.5163 [astro-ph.IM]

Koll\'ath, Z., Moln\'ar, L., Szab\'o, R., 2011, \textit{MNRAS}, 414, 1111

Kolenberg, K., et al., 2010, \textit{ApJ}, \textbf{713}, L198

Mas-Hesse, J.\ M., et al., 2003, \textit{A\&A}, \textbf{411,} L261

Mennickent, R.\ E., Gieren, W., Soszy\'nski, I., Pietrzy\'nski, G., 2006, \textit{A\&A}, \textbf{450}, 873

Miceli, A., et al., 2008, \textit{ApJ}, \textbf{678}, 865

Moln\'ar, L., Koll\'ath, Z., Szab\'o, R., Bryson, S., Kolenberg, K., Mullally, F., Thompson, S.\ E., 2012, \textit{ApJ}, \textbf{757}, L13

Moretti M.\ I., et al., 2009, \textit{ApJ}, \textbf{699}, L125

Moskalik, P. A., 2014, \textit{IAUS}, \textbf{301}, 249

Moskalik, P. A., Ko{\l}aczkowski, Z., 2009, \textit{MNRAS}, \textbf{394}, 1649

Neilson, H., Ignace, R., 2014, \textit{A\&A}, \textbf{563}, 4

Nemec, J.\ M., Nemec, A.\ F.\ L., Lutz, T.\ E., 1994, \textit{AJ}, \textbf{108}, 222

Nemec, J.\ M., Cohen, J.\ G., Ripepi, V., Derekas, A., Moskalik, P.\ A., Branimir, S., Chadid, M., Bruntt, H., 2013, \textit{MNRAS}, \textbf{773}, 181

Paunzen, E., Kuba, M., West, R.\ G., Zejda, M., 2014, \textit{IBVS}, No. 6090

Pietrzy\'nski, G., Gieren, W., Udalski, A., Bresolin, F., Kudritzki, R-P., Soszy\'nski, I., Szyma\'nski, M., Kubiak, M., 2004, \textit{AJ}, \textbf{128}, 2815

Plachy, E., Koll\'ath, Z., Moln\'ar, L., 2013, \textit{MNRAS}, \textbf{433}, 3590

Pojmanski, G., 1997, \textit{AcA}, \textbf{47}, 467

Rauer, H., et al., 2014, \textit{Exp. Astr.}, accepted, arXiv:1310.0696 [astro-ph.EP]

Samus, N.\ N., et al., 2004, Combined General Catalogue of Variable Stars

Schmidt, E.\ G., Reiswig, D.\ E., 1993, \textit{AJ}, \textbf{106}, 2429

Schmidt, E.\ G., et al., 2013, \textit{AJ}, \textbf{146}, 61

Sesar, B., et al., 2013, \textit{AJ}, \textbf{146}, 21

Smolec, R., 2005, \textit{AcA}, \textbf{55}, 59

Smolec, R., Moskalik, P.\ A., 2014, \textit{MNRAS}, \textbf{441}, 101

Smolec, R., et al., 2012, \textit{MNRAS}, \textbf{419}, 2407

Still, M., Barclay, T., 2012, Astrophysics Source Code Library, ascl:1208.004

Szabados, L., Kiss, L.\ L., Derekas, A., 2007, \textit{A\&A}, \textbf{461}, 613

Szab\'o, R., et al., 2011, \textit{MNRAS}, \textbf{413}, 2709

Szab\'o, R., et al., 2014, \textit{A\&A}, to be submitted

Wo\'zniak, P.\ R., et al., 2004, \textit{AJ}, \textbf{127}, 2436

\endreferences

\end{document}